\documentclass[preprint,review,12pt]{elsarticle}

\usepackage{graphics}

\usepackage{amssymb}
\usepackage{color}

\usepackage{lineno}

\journal{Icarus}

\begin{document}

\begin{frontmatter}




\title{The fate of ethane in Titan's hydrocarbon lakes and seas}

\author{Olivier~Mousis}
\ead{olivier.mousis@lam.fr}
\address{Aix Marseille Universit\'e, CNRS, LAM (Laboratoire d'Astrophysique de Marseille) UMR 7326, 13388, Marseille, France}
\address{Center for Radiophysics and Space Research, Space Sciences Building, Cornell University,  Ithaca, NY 14853, USA}
\author{Jonathan I. Lunine}
\address{Center for Radiophysics and Space Research, Space Sciences Building Cornell University,  Ithaca, NY 14853, USA}
\author{Alexander G. Hayes}
\address{Center for Radiophysics and Space Research, Space Sciences Building Cornell University,  Ithaca, NY 14853, USA}
\author{Jason D. Hofgartner}
\address{Department of Astronomy, Cornell University,  Ithaca, NY 14853, USA}

\begin{abstract}
Ethane is expected to be the dominant photochemical product on Titan's surface and, in the absence of a process that sequesters it from exposed surface reservoirs, a major constituent of its lakes and seas.  Absorption of Cassini's 2.2 cm radar by Ligeia Mare however suggests that this north polar sea is dominated by methane. In order to explain this apparent ethane deficiency, we explore the possibility that Ligeia Mare is the visible part of an alkanofer that interacted with an underlying clathrate layer and investigate the influence of this interaction on an assumed initial ethane-methane mixture in the liquid phase. We find that progressive liquid entrapment in clathrate allows the surface liquid reservoir to become methane-dominated for any initial ethane mole fraction below 0.75. If interactions between alkanofers and clathrates are common on Titan, this should lead to the emergence of many methane-dominated seas or lakes.
\end{abstract}


\end{frontmatter}


\section{Introduction}

Titan has a thick atmosphere dominated by nitrogen and methane. The dense orange-brown smog hiding the satellite's surface is produced by photochemical reactions of methane, nitrogen, and their dissociation products with solar ultraviolet, which lead primarily to the formation of ethane and heavier hydrocarbons (Lavvas et al., 2008a, 2008b). Ethane and propane are expected to accumulate in the surface liquids and thus to be important constituents of liquid bodies that are in equilibrium with the atmosphere (Cordier et al., 2009; Cordier et al., 2013; Tan et al., 2013). This classical picture, derived from Titan's photochemistry models, {contrasts} with the indirect measurements of the composition of Ligeia Mare obtained via radar absorption, suggesting that this sea is dominated by methane (Mastrogiuseppe et al., 2014; Mitchell et al., 2015). {There is thus a discrepancy between the measured ethane/methane ratio and the expectation from otherwise well-tested photochemical schemes for Titan's atmospheric methane.}

On the other hand, the temperature and atmospheric pressure conditions prevailing at Titan's ground level permit clathrate formation when liquid hydrocarbons enter in contact with the exposed water ice (Mousis and Schmitt, 2008). Assuming a high porosity for Titan's upper crust, clathrates with hydrocarbon guest species are stable and expected to occur down to several kilometers from the surface (Mousis and Schmitt, 2008; Choukroun and Sotin, 2012). These clathrates may contain a significant fraction of the ethane and propane generated in Titan's atmosphere over the solar system's lifetime (Mousis and Schmitt, 2008). Also, under some circumstances, the lakes and seas present on Titan (Stofan et al., 2007; Mastrogiuseppe et al., 2014) can interact with underlying clathrate layers and, as a result, present compositions different from those expected from atmosphere--sea equilibrium (Mousis et al., 2014). In particular, if the basins containing liquids are in contact with the icy crust of Titan, their compositions may be altered by interactions with clathrate reservoirs that progressively form if the liquid mixtures diffuse throughout preexisting porous icy layers (Mousis et al., 2014). 

{The model proposed by Mousis et al. (2014) predicted that, if Titan's lakes interacted with clathrate reservoirs, they could become dominated by ethane and/or propane and be extremely impoverished in methane. This conclusion was based on an initial lake composition equal to that from the model of Cordier et al. (2009, 2013). However, several recent studies point toward very different compositions of the lakes and seas, which depend on the {particular} thermodynamic approach (Glein and Shock, 2013; Tan et al., 2013; Tan et al., 2015; Luspay-Kuti et al., 2015). The initial lake composition is a key parameter because it determines the amounts of the different species that become entrapped in clathrate and thus the ultimate composition of the lakes and seas.

Here, we utilize the model proposed by Mousis et al. (2014) to investigate the range of initial compositions of Titan's Ligeia Mare that could be consistent with the recent measurements that {show it to be} dominated by methane (Mastrogiuseppe et al., 2014; Mitchell et al., 2015). We {demonstrate} that a sea will become methane-dominated for any initial ethane mole fraction below 0.75 if it interacted with an underlying clathrate layer that progressively formed from the entrapment of methane and ethane.} 


\section{Model description}

Our liquid reservoir is considered as a mixture of methane and ethane, and its equilibration is assumed faster with clathrate than with the atmosphere. This implies that the considered hydrocarbon lake or sea is the only visible part of a larger alkanofer that is in contact with the icy porous crust. We follow the approach proposed by Mousis et al. (2014), who consider an isolated system composed of a clathrate reservoir that progressively forms and replaces the crustal material with time and a {well-mixed liquid reservoir} that correspondingly empties due to the net transfer of molecules to the clathrate reservoir. We use the numerical procedure defined in Mousis et al.  (2014) with the intent to determine the mole fractions of each species present in the liquid reservoir and trapped in the forming clathrate reservoir. {These mole fractions depend on the species initial fractions (before volatile migration) in the liquid and clathrate.} Our computations start from a predefined composition of the liquid reservoir. They use an iterative process for which the number of moles in the liquid phase being trapped in clathrates between each iteration is equal to 10$^{-4 }$ the total number of moles available. The numerical procedure utilized to calculate at each step the relative abundances of guest species incorporated in clathrates is based on a statistical mechanics model that relates their macroscopic thermodynamic properties to the molecular structures and interaction energies (van der Waals and Platteeuw, 1959; Lunine and Stevenson, 1985; Sloan and Koh, 2008; Mousis et al. 2014).
	
In this approach, the fractional occupancy of a guest molecule $K$ for a given type $q$ ($q$~=~small or large; see Sloan and Koh, 2008) of cage is written as

\begin{equation}
y_{K,q}=\frac{C_{K,q}f_K}{1+\sum_{J}C_{J,q}f_J} ,
\label{eq1}
\end{equation}

\noindent where the sum in the denominator includes all the species which are present in the liquid phase (here methane and ethane). $C_{K,q}$ is the Langmuir constant of species $K$ in the cage of type $q$, and $f_K$  the fugacity of species $K$ in the mixture. Using the Redlich-Kwong equation of state (Redlich and Kwong, 1949) in the case of a mixture dominated either by methane or ethane, we find that its coefficient of fugacity $\phi$ converges towards 1 at Titan's surface conditions, implying that $f_K$ converges towards $P_K$, namely the vapor pressure of species $K$. 

{We consider the liquid hydrocarbon mixture as an ideal solution and the value $f_K$ of each species $K$ can then be calculated via the Raoult's law, which states}

\begin{equation}
f_K \simeq  P_K = x^{lake}_K \times P^*_K,
\label{eq2}
\end{equation}

\noindent with $P^*_K$  the vapor pressure of pure component $K$ in the liquid, defined via the Antoine equation given in Mousis et al. (2014) and $x^{lake}_K$ the mole fraction of pure component $K$ in the liquid.

The Langmuir constant, which depends on the strength of the interaction between each guest species and each type of cage, is determined by integrating the molecular potential within the cavity as

\begin{equation}
C_{K,q}=\frac{4\pi}{k_B T}\int_{0}^{R_c}\exp\Big(-\frac{w_{K,q}(r)}{k_B T}\Big)r^2dr ,
\label{eq3}
\end{equation}

\noindent where $R_c$ represents the radius of the cavity assumed to be spherical, $k_B$ the Boltzmann constant, and $w_{K,q}(r)$ is the spherically averaged Kihara potential representing the interactions between the guest molecules $K$ and the H$_2$O molecules forming the surrounding cage $q$. In our formalism, $w(r)$ is written for a spherical guest molecule (McKoy and Sinano\u{g}lu, 1963):

\begin{eqnarray}
w(r) 	= 2z\epsilon\Big[\frac{\sigma^{12}}{R_c^{11}r}\Big(\delta^{10}(r)+\frac{a}{R_c}\delta^{11}(r)\Big) \\ \nonumber
- \frac{\sigma^6}{R_c^5r}\Big(\delta^4(r)+\frac{a}{R_c}\delta^5(r)\Big)\Big],
\label{eq4}
\end{eqnarray}

\noindent with

\begin{eqnarray}
\delta^N(r)=\frac{1}{N}\Big[\Big(1-\frac{r}{R_c}-\frac{a}{R_c}\Big)^{-N}- \\ \nonumber
\Big(1+\frac{r}{R_c}-\frac{a}{R_c}\Big)^{-N}\Big].
\label{eq5}
\end{eqnarray}

\noindent In Eq. 4, $z$ is the coordination number of the cell. Its value, taken from Sloan and Koh (2008), depends on the clathrate structure (I or II) and on the type of the cage (small or large). The Kihara parameters $a$, $\sigma$ and $\epsilon$ for the molecule-water interactions are also derived from Sloan and Koh (2008).

Finally, the mole fraction $x^{clat}_K$ of a guest molecule $K$ in a clathrate can be calculated with respect to the whole set of species considered in the system as

\begin{equation}
x^{clat}_K=\frac{b_s y_{K,s}+b_\ell y_{K,\ell}}{b_s \sum_J{y_{J,s}}+b_\ell \sum_J{y_{J,\ell}}},
\label{eq5}
\end{equation}

\noindent where $b_s$ and $b_l$ are the number of small and large cages per unit cell respectively, for the clathrate structure under consideration, and with $\sum_K$~$x^{clat}_K~=~1$. Values of $R_c$, $z$, $b_s$ and $b_l$ are taken from Sloan and Koh (2008). All the computations performed here are based on the assumption that only structure I clathrates form at equilibrium from a mixture of methane and ethane, as suggested by experiments Takeya et al. (2003).


\section{Results}

Figure \ref{plot1} represents the evolution of the mole fractions of species present in the liquid reservoir and its associated clathrate as a function of the progressive liquid entrapment. Our calculations have been conducted at a surface temperature of 91 K and the starting mole fractions of methane and ethane have been set to 0.7 and 0.3 in the liquid reservoir, respectively. These mole fractions are close to the values found in the lake-atmosphere model of Tan et al (2013). The mole fraction of ethane in the liquid significantly decreases with progressive entrapment of the liquid reservoir, forcing this reservoir to become methane-pure when more than $\sim$half (in mole fraction) of the initial reservoir has been trapped in clathrates. In this case, the mole fraction of ethane is below 5\% in the liquid phase when more than $\sim$40\% of the initial liquid reservoir is trapped in clathrate. The significant sink of ethane in the liquid phase results from its more efficient trapping in clathrate compared to methane, as illustrated by Fig. \ref{plot1} at low mole fractions of entrapped liquid. As the liquid reservoir becomes ethane depleted, progressive liquid entrapment occurs primarily for methane molecules and the mole fraction of methane in the clathrate increases. At the end of the liquid reservoir enclathration, the mole fractions of trapped methane and ethane converge towards their starting values in the liquid.

Figure \ref{plot2} displays the starting mole fractions of ethane in the liquid phase and corresponding fractions of the remaining liquid reservoir that give methane mole fractions of 0.90, 0.95 and 0.99. {Our computations have been performed at a surface temperature of 91 K but a temperature difference of a few K would have not affected the results. The Figure} shows that ethane can be severely depleted in the liquid reservoir for a wide range of starting mole fractions. All starting mole fractions converge towards 0.75 when the fraction of remaining liquid is close to zero. The value of 0.75 is the ratio of the number of large cages to the total number of cages for structure I clathrate. Ethane can only be trapped in the large cages and thus the ethane mole fraction never exceeds 0.75 in clathrate (Sloan and Koh, 2008). Thus for a starting composition near an ethane mole fraction of 0.75 nearly all of the liquid must be enclathrated before {there} is an appreciable change in the liquid composition.  This explains why all of the curves converge toward this value when the fraction of remaining liquid is zero.

\section{Discussion}

The work of Mastrogiuseppe et al. (2014) indicates that Ligeia Mare in the north is dominated by methane with ethane considered at least as a secondary component. Cassini Visual and Infrared Mapping Spectrometer data indicate the presence of liquid ethane in Ontario Lacus (Brown et al., 2008) but this could be consistent with small ethane mixing ratios ($\sim$1\%). Circumstantial evidence from a variety of geologic features and isotopic ratios of carbon, hydrogen, and potassium in the atmosphere suggest that methane resupply from the interior has been extensive, enough to produce vastly more ethane than is seen on the surface (Lorenz et al., 2008). Our model shows that the ethane will then naturally be incorporated in the clathrate. One issue is how to ``renew'' the crustal water ice once saturated with ethane clathrate; the subsidence model of Choukroun and Sotin (2012) seems promising in this regard. 

{A key assumption in our model is that the liquid reservoirs equilibrate faster with clathrates than with the atmosphere. Recent experiments suggest that the timescales involved for methane clathrate formation and ethane substitution of methane in clathrate are of the order of a few hours at moderate pressures (30--40 bar) and temperatures in the 220--250 K range (Vu and Choukroun, 2015). To be relevant with Titan's subsurface conditions, measurements of clathrate formation kinetics still have to be performed in the 90--100 K range. On the other hand, the timescales for sea-atmosphere equilibrium on Titan are also not well understood due to a lack of constraints on the kinetics at the relevant conditions.  Also, the surface liquids and atmosphere may not be in equilibrium because dynamic processes (such as rain) may be operating on timescales shorter than those required for the lakes and seas to equilibrate with the atmosphere (Sotin et al., 2012). Regardless, given the substantial depth of the sea (Mastrogiuseppe et al., 2014), it must be recognized that the composition potentially can vary between the top and bottom of the sea as clathration is occurring, and ethane is drawn out of the liquid at the lake bottom. However, the lower density of methane compared to ethane ensures that the methane-rich seabottom liquid will rise upward, and the more ethane-rich overlying fluid mix downward, so that in reality we expect the sea to remain compositionally homogeneous as clathration proceeds.}

Other hypotheses to explain the ethane deficiency in Ligeia Mare have been presented. Lorenz et al. (2014) for example suggested that Ligeia Mare is not in equilibrium with the atmosphere because it is sourced by methane precipitation and exports its ethane to Kraken Mare on a timescale faster than atmosphere--sea equilibrium. This mechanism predicts a difference in the ethane/methane ratio between Ligeia and Kraken, but is not sufficient to explain a more global ethane deficiency in the lakes or seas of Titan.

Our results make intuitive sense when one considers that in the presence of abundant methane and ethane, clathrate will form with an ethane mole fraction of 0.75.  From this consideration we can deduce that any liquid reservoir with an initial ethane mole fraction of less than 0.75 will become more methane rich as a result of liquid enclathration. Similary a liquid reservoir with an initial ethane mole fraction of greater than 0.75 will become more ethane rich as a result of liquid enclathration. Therefore we can set 0.75 as the upper limit on the amount of ethane that was initially present in the liquid reservoir if the current composition of Ligeia Mare is the result of hydrocarbon trapping in a coexisting clathrate layer. {One consequence of the progressive methane enrichment in the liquid is the corresponding increase of its vapor pressure. Thus the increased methane abundance in the liquid for lakes and seas that have a progressively forming clathrate layer may lead to increased methane rainfall near these bodies.

Simulations representing clathrate formation from a ternary methane-ethane-nitrogen system do not change our results. Because N$_2$ is poorly trapped by clathrates, the ternary system would progressively become a binary methane-nitrogen system with a maximum mole fraction of $\sim$20$\%$ for nitrogen in the liquid at 90 K (value corresponding to its solubility limit in hydrocarbons; Hibbard and Evans, 1968).}

It is quite likely that Titan's water ice crust has been largely methane clathrate over its history (Tobie et al., 2006). In this case, our model is unchanged, because ethane's preferential incorporation in clathrate relative to methane means that the latter is driven out of the ice as the former is incorporated, {provided that the energy barrier for this substitution is not too high (Vu and Choukroun, 2015). Because two methane molecules need to be photochemically destroyed to produce one ethane molecule, the substitution mechanism alone cannot explain the replenishment of Titan's atmospheric methane over its evolution. Alternative mechanims such as impacts and cryovolcanic events {may be} needed to explain the present methane in Titan's atmosphere (Chroukroun and Sotin, 2012). Locally,} a signature of such mediation of the exchange of ethane and methane over time might be heterogeneity of the isotopic signatures ($^{12}$C/$^{13}$C and D/H) of methane in the seas over fracture zones where the methane is outgassing. A heterogeneity of the isotopic composition of methane should be also observed between the sea and the atmosphere {if the sea is being actively} supplied with crustal methane. Either way, testing this requires a sea lander (Stofan et al., 2013).

Could the ethane deficiency on Titan be due to the interactions between lakes-seas and clathrate layers? Global circulation models predict that the downwelling circulation branch surrounding the poles during the winter induces ethane condensation in the tropopause and its precipitation to the surface (Griffith et al., 2006; Rannou et al., 2006; Mitchell et al., 2009). Assuming that the {extant} atmospheric methane has been outgassing at the current rate for only 0.6 gigayears (Gyr) (Tobie et al., 2006) and that the ethane production rate is 4.1 $\times$ 10$^9$ cm$^{-2}$ s$^{-1}$ (170 kg s$^{-1}$) (Lavvas et al., 2008a, 2008b), the amount of produced ethane should be sufficient to form an ocean with a globally averaged thickness of $\sim$70 m on Titan. If all this ethane has been trapped in the large cages of structure I clathrates by replacing the methane molecules following the mechanism proposed by Choukroun and Sotin (2012), the corresponding thickness of the clathrate layer should be of $\sim$221 m globally averaged on the satellite. If the produced ethane has been brought via fluvial transport to its lakes and seas, then {the latter} must have been in contact with a clathrate reservoir whose equivalent area is $\sim$7 times the one estimated for the lakes and seas {($\sim$1.1\% of the satellite's surface; Hayes et al., 2008)} to have a thickness of $\sim$2900 m, a value corresponding to {that} imposed by the subsidence mechanism proposed by Choukroun and Sotin (2012), and in any event much less than the 40 km and 55--80 km crustal thicknesses determined by Iess et al (2012) and Beghin et al. (2012), respectively.

\section*{Acknowledgments}

O.M. acknowledges support from CNES. J.I.L. acknowledges support from the Cassini project. J.D.H. acknowledges support from NSERC. The work contributed by O.M. was carried out thanks to the support of the A*MIDEX project (n\textsuperscript{o} ANR-11-IDEX-0001-02) funded by the ``Investissements d'Avenir'' French Government program, managed by the French National Research Agency (ANR).


\begin{thebibliography}{00}

\bibitem[B{\'e}ghin et al.(2012)]{2012Icar..218.1028B} B{\'e}ghin, C., Randriamboarison, O., Hamelin, M., Karkoschka, E., Sotin, C., Whitten, R.~C., Berthelier, J.-J., Grard, R., Sim{\~o}es, F.\ 2012.\ Analytic theory of Titan's Schumann resonance: Constraints on ionospheric conductivity and buried water ocean.\ Icarus 218, 1028-1042.

\bibitem[Brown et al.(2008)]{B08} Brown, R.~H., Soderblom, L.~A., Soderblom, J.~M., Clark, R.~N., Jaumann, R., Barnes, J.~W., Sotin, C., Buratti, B., Baines, K.~H., Nicholson, P.~D.\ 2008.\ The identification of liquid ethane in Titan's Ontario Lacus.\ Nature 454, 607-610. 

\bibitem[Choukroun and Sotin(2012)]{CS12} Choukroun, M., Sotin, C.\ 2012.\ Is Titan's shape caused by its meteorology and carbon cycle?\ Geophysical Research Letters 39, 4201.

\bibitem[Cordier et al.(2013)]{2013ApJ...768L..23C} Cordier, D., Mousis, O., Lunine, J.~I., Lavvas, P., Vuitton, V.\ 2013.\ Erratum: ``An Estimate of the Chemical Composition of Titan's Lakes''. The Astrophysical Journal 768, L23.

\bibitem[Cordier et al.(2009)]{C09} Cordier, D., Mousis, O., Lunine, J.~I., Lavvas, P., Vuitton, V.\ 2009.\ An Estimate of the Chemical Composition of Titan's Lakes.\ The Astrophysical Journal 707, L128-L131.

\bibitem[Glein and Shock(2013)]{2013GeCoA.115..217G} Glein, C.~R., Shock, E.~L.\ 2013.\ A geochemical model of non-ideal solutions in the methane-ethane-propane-nitrogen-acetylene system on Titan.\ Geochimica et Cosmochimica Acta 115, 217-240.

\bibitem[Griffith et al.(2006)]{G06} Griffith, C.~A., and 13 colleagues 2006.\ Evidence for a Polar Ethane Cloud on Titan.\ Science 313, 1620-1622. 

\bibitem[Hayes et al.(2008)]{H08} Hayes, A., and 13 colleagues 2008.\ Hydrocarbon lakes on Titan: Distribution and interaction with a porous regolith.\ Geophysical Research Letters 35, L09204. 

\bibitem[Hibbard and Stevenson(1985)]{Hibbard...58..493L} Hibbard, R. R., Evans, Jr. A. 1968. On the solubilities and rates of solution of gases in liquid methane. NASA Technical Note, NASA TN, D-4701.

\bibitem[Iess et al.(2012)]{2012Sci...337..457I} Iess, L., Jacobson, R.~A., Ducci, M., Stevenson, D.~J., Lunine, J.~I., Armstrong, J.~W., Asmar, S.~W., Racioppa, P., Rappaport, N.~J., Tortora, P.\ 2012.\ The Tides of Titan.\ Science 337, 457.

\bibitem[Lavvas et al.(2008)]{Lavvas08a} Lavvas, P.~P., Coustenis, A., Vardavas, I.~M.\ 2008.\ Coupling photochemistry with haze formation in Titan's atmosphere, Part I: Model description.\ Planetary and Space Science 56, 27-66. 

\bibitem[Lavvas et al.(2008)]{Lavvas08b} Lavvas, P.~P., Coustenis, A., Vardavas, I.~M.\ 2008.\ Coupling photochemistry with haze formation in Titan's atmosphere, Part II: Results and validation with Cassini/Huygens data.\ Planetary and Space Science 56, 67-99. 

\bibitem[Lorenz(2014)]{Lo14} Lorenz, R.~D.\ 2014.\ The flushing of Ligeia: Composition variations across Titan's seas in a simple hydrological model.\ Geophysical Research Letters 41, 5764-5770. 

\bibitem[Lorenz et al.(2008)]{L08} Lorenz, R.~D., and 15 colleagues 2008.\ Titan's inventory of organic surface materials.\ Geophysical Research Letters 35, L02206. 

\bibitem[Lunine and Stevenson(1985)]{LS85} Lunine, J.~I., Stevenson, D.~J.\ 1985.\ Thermodynamics of clathrate hydrate at low and high pressures with application to the outer solar system.\ The Astrophysical Journal Supplement Series 58, 493-531. 

\bibitem[Luspay-Kuti et al.(2015)]{2015E&PSL.410...75L} Luspay-Kuti, A., Chevrier, V.~F., Cordier, D., Rivera-Valentin, E.~G., Singh, S., Wagner, A., Wasiak, F.~C.\ 2015.\ Experimental constraints on the composition and dynamics of Titan's polar lakes.\ Earth and Planetary Science Letters 410, 75-83.

\bibitem[Mastrogiuseppe et al.(2014)]{Mas14} Mastrogiuseppe, M., and 11 colleagues 2014.\ The bathymetry of a Titan sea.\ Geophysical Research Letters 41, 1432-1437.

\bibitem[McKoy \& Sinano\u{g}lu (1963)]{MS63} McKoy, V., Sinano\u{g}lu, O., 1963.\ Theory of dissociation pressures of some gas hydrates.\ Journal of Chemical Physics 38 (12), 2946-2956.

\bibitem[Mitchell et al.(2015)]{Mit15} Mitchell, K.~L., Barmatz, M., Jamieson, C. S.\ 2015.\ Laboratory Measurements of Cryogenic Liquid Alkane Microwave Absorptivity and Implications for the Composition of Ligeia Mare, Titan.\ Geophys. Res. Lett., in press. 

\bibitem[Mitchell et al.(2009)]{Mit09} Mitchell, J.~L., Pierrehumbert, R.~T., Frierson, D.~M.~W., Caballero, R.\ 2009.\ The impact of methane thermodynamics on seasonal convection and circulation in a model Titan atmosphere.\ Icarus 203, 250-264. 

\bibitem[Mousis et al.(2014)]{Mousis14} Mousis, O., Choukroun, M., Lunine, J.~I., Sotin, C.\ 2014.\ Equilibrium composition between liquid and clathrate reservoirs on Titan.\ Icarus 239, 39-45. 

\bibitem[Mousis and Schmitt(2008)]{MS08} Mousis, O., Schmitt, B.\ 2008.\ Sequestration of Ethane in the Cryovolcanic Subsurface of Titan.\ The Astrophysical Journal 677, L67-L70.


\bibitem[Rannou et al.(2006)]{R06} Rannou, P., Montmessin, F., Hourdin, F., Lebonnois, S.\ 2006.\ The Latitudinal Distribution of Clouds on Titan.\ Science 311, 201-205. 

\bibitem[Redlich et al.(1949)]{RK49} Redlich, O., Kwong, J. N. S. 1949. On the Thermodynamics of Solutions. V. An Equation of State. Fugacities of Gaseous Solutions. Chemical Reviews 44 (1): 233-244.

\bibitem[Sloan et al. (2008)]{Sloan08} Sloan, E. D., Koh, C. A., 2008. Clathrate Hydrates of Natural Gases. 3rd ed.; CRC Press, Taylor \& Francis Group, Boca Raton.

\bibitem[Stofan et al.(2013)]{S13} Stofan, E., Lorenz, R., Lunine, J., Bierhaus, E.~B., Clark, B., Mahaffy, P.~R., Ravine, M.\ 2013.\ TiME - The Titan Mare Explorer.\ Proceedings of the 2013 IEEE Aerospace Conference 211. 

\bibitem[Stofan et al.(2007)]{Sto07} Stofan, E.~R., and 37 colleagues 2007.\ The lakes of Titan.\ Nature 445, 61-64. 

\bibitem[Takeya et al.(2003)]{Takeya03} Takeya, S., Kamata, Y., Uchida, T., Nagao, J., Ebinuma, T., Narita, H., Hori, A., and Hondoh, T., 2003. Coexistence of structure I and II hydrates formed from a mixture of methane and ethane gases. Can. J. Phys. 81, 479-484.

\bibitem[Tan et al.(2015)]{2015Icar..250...64T} Tan, S.~P., Kargel, J.~S., Jennings, D.~E., Mastrogiuseppe, M., Adidharma, H., Marion, G.~M.\ 2015.\ Titan's liquids: Exotic behavior and its implications on global fluid circulation.\ Icarus 250, 64-75.

\bibitem[Tan et al.(2013)]{2013Icar..222...53T} Tan, S.~P., Kargel, J.~S., Marion, G.~M.\ 2013.\ Titan's atmosphere and surface liquid: New calculation using Statistical Associating Fluid Theory.\ Icarus 222, 53-72.

\bibitem[Tan et al.(2013)]{T13} Tan, S.~P., Kargel, J.~S., Marion, G.~M.\ 2013.\ Titan's atmosphere and surface liquid: New calculation using Statistical Associating Fluid Theory.\ Icarus 222, 53-72. 

\bibitem[Tobie et al.(2006)]{T06} Tobie, G., Lunine, J.~I., Sotin, C.\ 2006.\ Episodic outgassing as the origin of atmospheric methane on Titan.\ Nature 440, 61-64.

\bibitem[van der Waals \& Platteeuw (1959)]{vdWP59} van der Waals, J.~H., Platteeuw, J.~C., 1959.\ Clathrate solutions. In: Advances in Chemical Physics, Vol. 2, Interscience, New York, 1-57.

\bibitem[Vu and Choukroun(2015)]{2015LPI....46.2484V} Vu, T.~H., Choukroun, M.\ 2015.\ Experimental Studies of Methane Clathrate Formation and Substitution with Ethane.\ Lunar and Planetary Science Conference 46, 2484.

\clearpage

\begin{figure*}[h]
\begin{center}
\resizebox{\hsize}{!}{\includegraphics[angle=0]{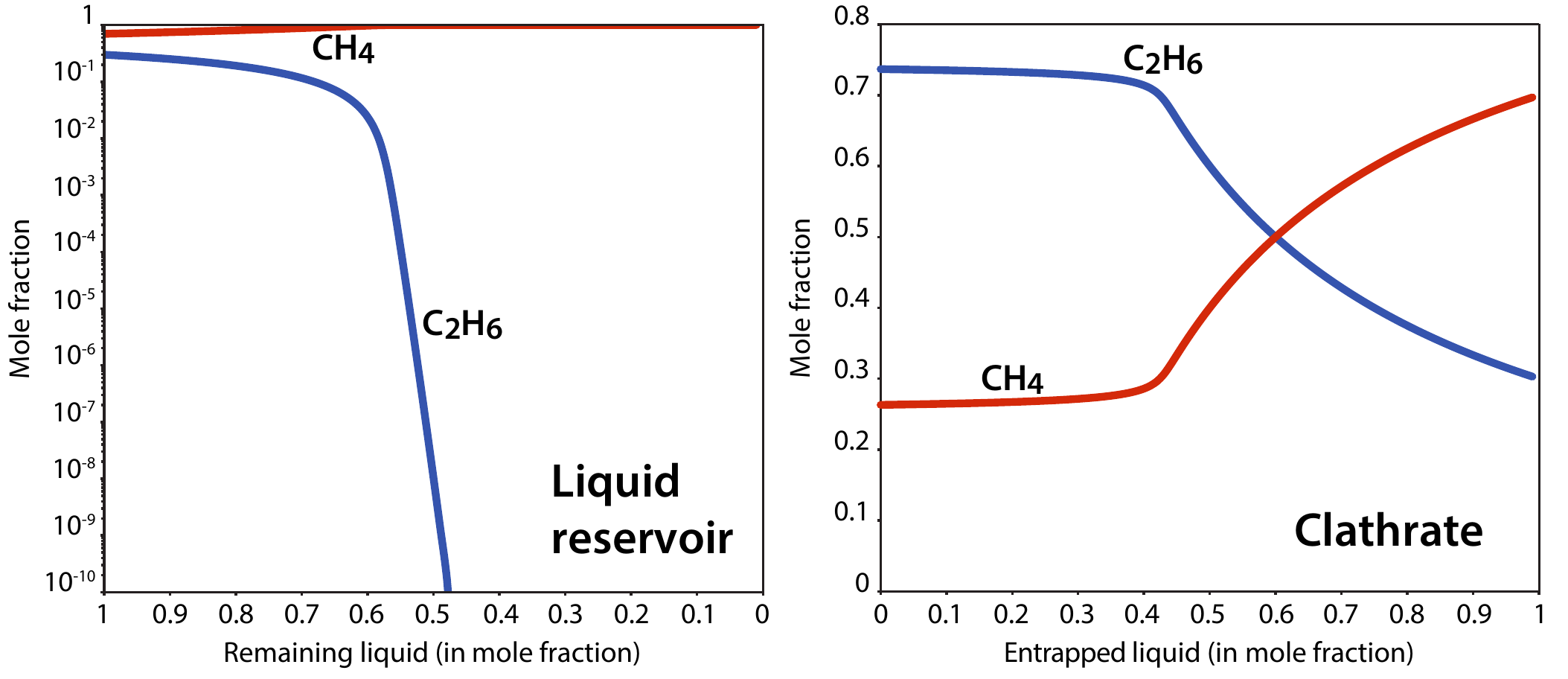}}
\caption{Left: composition of a liquid reservoir in contact with forming clathrate. Mole fractions of methane and ethane are expressed as a function of the remaining fraction of the initial liquid reservoir. Right: corresponding composition of clathrate in contact with the liquid reservoir. Mole fractions of methane and ethane are expressed as a function of the mole fraction of the initial liquid reservoir that is entrapped.}
\label{plot1}
\end{center}
\end{figure*}

\clearpage

\begin{figure}[h]
\begin{center}
\resizebox{\hsize}{!}{\includegraphics[angle=0]{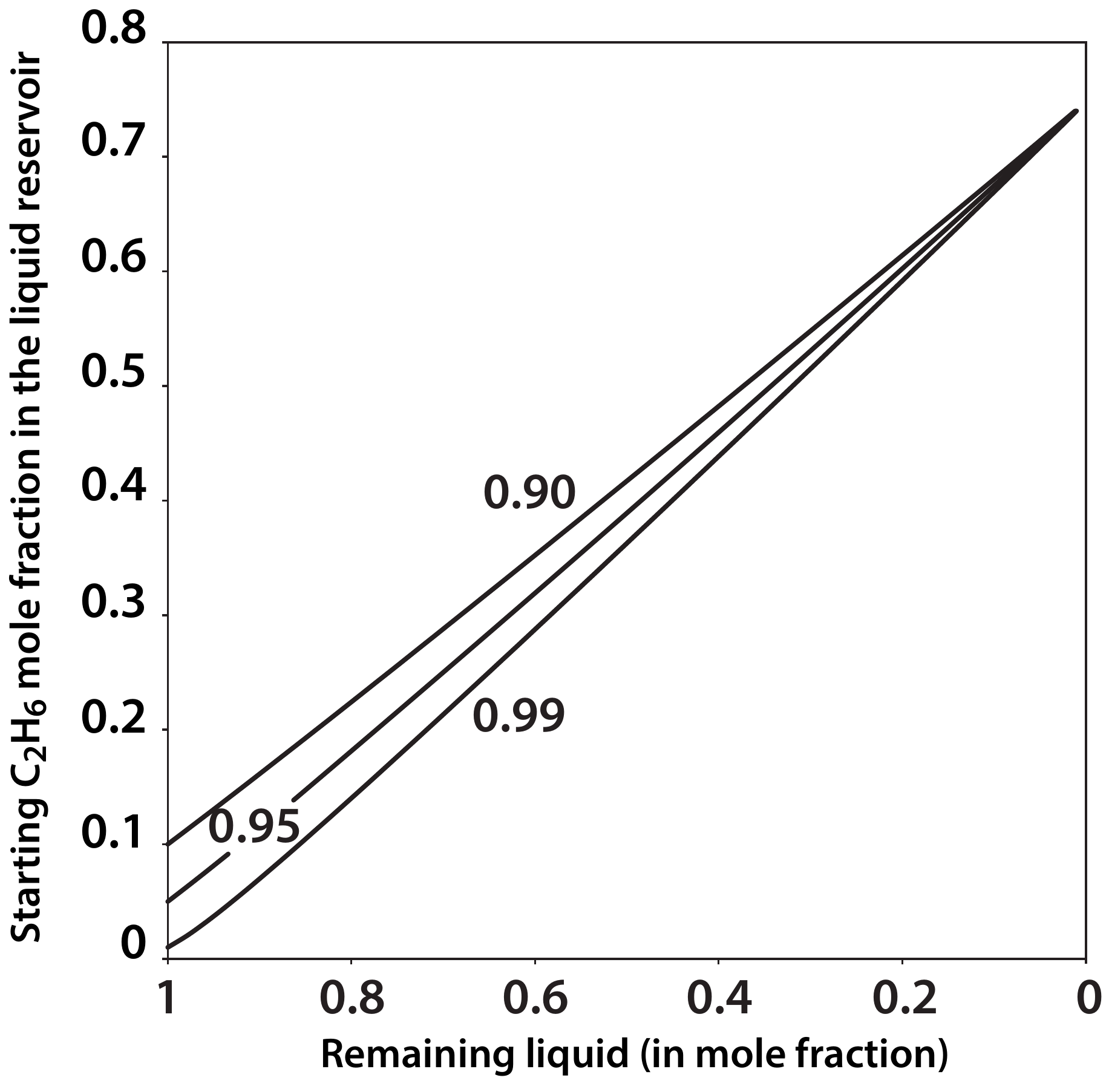}}
\caption{Starting mole fractions of ethane in the liquid phase and corresponding fractions of the remaining liquid reservoir that give methane mole fractions of 0.90, 0.95 and 0.99.}
\label{plot2}
\end{center}
\end{figure}










\end{thebibliography}
\end{document}